\documentclass{mn2e}
\input{epsf}

\def \vt{\vartheta}
\def \farcs{\hbox{$.\!\!^{\prime\prime}$}}

\title[Imperfect correction of PSF anisotropy]
{The effect of imperfect corrections of PSF anisotropy on
cosmic shear measurements$^1$}

\author[Henk Hoekstra]{Henk Hoekstra\\
Canadian Institute for Theoretical Astrophysics,
University of Toronto, Toronto, M5S 3H8, Canada\\
Department of Astronomy and Astrophysics,
University of Toronto, Toronto, M5S 3H8, Canada}

\begin{document}

\maketitle

\begin{abstract} 

Current measurements of the weak lensing signal induced by large scale
structure provide useful constraints on a range of cosmological
parameters. However, the ultimate succes of this technique depends on
the accuracy with which one can correct for the effect of the Point
Spread Function (PSF), in particular the correction for the PSF
anisotropy. With upcoming large weak lensing surveys a proper
understanding of residual systematics is necessary.

In this paper we examine the accuracy of the PSF anisotropy correction
using images of fields with a large number of stars. A randomly
selected subset of stars is used to characterize the PSF and to
correct the shapes of the remaining stars. The ellipticity correlation
function of the residuals is studied to quantify the effect of
imperfect corrections for PSF anisotropy on cosmic shear
studies. These imperfections occur on the chip scale and consequently
the systematic signal decreases rapidly with increasing angular
scale. Separation of the signal into ``E'' (curl-free) and ``B''
(curl) components can help to identify the presence of residual
systematics, but in general, the amplitude of the ``B''-mode is
different from that of the ``E''-mode.

The study of fields with many stars can be beneficial in finding a
proper description of the variation of PSF anisotropy, and
consequently help to significantly improve the accuracy with which the
cosmic shear signal can be measured.  We show that with such an
approach it is feasible that the accuracy of future cosmic shear
studies is limited by the statistical noise introduced by the
intrinsic shapes of the sources. In particular, the prospects for
accurate measurements of the cosmic shear signal on scales larger than
$\sim 10$ arcminutes are excellent.

\end{abstract}


\section{Introduction}
\footnotetext[1]{Based on observations from the Canada-France-Hawaii
Telescope, which is operated by the National Research Council of
Canada, le Centre Nationale de la Recherche Scientifique and the
University of Hawaii.}

Intervening large scale structure causes a systematic distortion in
the images of distant galaxies. The amplitude of this effect is small,
but measurable, and provides a direct measure of the clustering of
matter in the universe. In recent years several groups have reported
measurements of this cosmic shear signal (e.g., Bacon et al. 2000,
2003; Brown et al. 2003; Hamana et al. 2002; Hoekstra et
al. 2002a,2002b; Jarvis et al. 2003; Kaiser et al. 2000; Maoli et
al. 2001; Refregier et al. 2002; van Waerbeke et al. 2000, 2001, 2002;
Wittman et al. 2000).

Already these measurements can be used to constrain cosmological
parameters, in particular when combined with CMB measurements (e.g.,
Contaldi, Hoekstra \& Lewis 2003). Several much larger weak lensing
studies are currently underway or will start in the near future. In
particular, the Canada-France-Hawaii Telescope Legacy Survey (CFHTLS)
will be a major step forward. The survey aims to image $\sim 168$
square degrees in five filters, down to $I_{\rm AB}=24.5$. The
multi-colour imaging data allow for photometric redshifts to be
determined, which enables us to measure the redshift evolution of the
matter power spectrum and remove the contribution from intrinsic
alignements of the sources.

Thus weak lensing has the prospects of becoming one of the key methods
to be used in observational cosmology. However, the accuracy with
which the cosmic shear signal can be measured depends critically on
the correction for the effects of the Point Spread Function (PSF).

The PSF corrupts the shapes of the galaxies used to measure the weak
lensing signal. The circularisation of the images by the PSF (seeing)
systematically lowers the lensing signal. A correction is needed to
relate the observed ellipticities to the true shapes (e.g., Luppino \&
Kaiser 1997; Hoekstra et al. 1998). In addition, the PSF typically is
anisotropic which results in a coherent distortion of the shapes of
the sources, mimicking a weak lensing signal. The PSF anisotropy is
typically comparable or larger than the weak lensing signal one
intends to measure.

Several techniques have been developed to correct the observed shapes
of faint galaxies (e.g., Kaiser, Squires, \& Broadhurst 1995; Luppino
\& Kaiser 1997, Hoekstra et al. 1998; Kuijken 1999; Bernstein \&
Jarvis 2002; Hirata \& Seljak 2003). In particular the method proposed
by Kaiser et al. (1995) is widely used and has been tested extensively
(Hoekstra et al. 1998; Erben et al. 2001; Bacon et al. 2001). These
studies suggest that the corrections work rather well, but more work
is required to ensure the success of future surveys.

Although much work has been devoted to improving the shape
measurements and better correction schemes, one of the key elements in
the correction process has been ignored: how well can one characterize
the spatial variation of the PSF anisotropy? The use of an incorrect
model will result in a residual signal, no matter how sophisticated
the correction algorithm is. 

In this paper we examine this problem using {\it real} data and
quantify the impact on cosmic shear measurements. In actual weak
lensing studies the correction scheme might leave systematic residuals
that depend on the size or profile of the galaxies. Here we only
consider stars, and consequently the correction for PSF anisotropy is
ideal (apart from noise present in the data).

Most current weak lensing studies use data from mosaic cameras, which
have $8-32$ 2k$\times$4k chips. The PSF changes from chip to chip, and
it is therefore not possible to fit a model of the PSF anisotropy to
the whole mosaic. In practice one derives a model for each chip
separately. For each individual chip, about $50-100$ stars can be used
to measure the PSF anisotropy as a function of position. We examine
whether such a limited number of stars sufficient to characterize the
variation of PSF anisotropy.

The structure of the paper is as follows. In \S2 we discuss the shape
measurements and the correction for PSF anisotropy. In \S3 we examine
how well standard correction schemes, which typically fit a second
order polynomial model to the PSF shape measurements, perform. We also
discuss strategies which can reduce the residual signal. We separate
the measurements into ``E'' and ``B'' modes and compare the signals.
In \S4 we quantify how imperfect PSF anisotropy corrections will
affect weak lensing measurements. We also examine the influence of seeing
on the accuracy of cosmic shear measurements.

\section{Analysis}

We use data taken with the CFHT using the CFH12k camera. The data were
obtained as part of the EXPLORE project (Mall{\'e}n-Ornelas 2003; Yee
et al.  2003) which aims to find planets transiting stars. To maximize
the probability of finding planets, the fields contain a large number
of stars, which make the data ideal for the study presented here.
Furthermore the same field is followed during the whole night in order
to sample the light curves over a long period of time. As a result we
can also examine how the PSF anisotropy changes with time, in
particular when the orientation of the telescope is completely
different.

Table~\ref{tab_data} lists the CFHT exposure numbers, modified Julian
dates of the exposures and exposure times, as well as the seeing. All
exposures were taken in the $R$-band. We included a series of four
exposures taken within a 10 minute period to study the stability of
the PSF pattern over short periods of time. The stability over longer
periods of time is examined by including an image taken the next
night.

\begin{table}
\begin{center}
\begin{tabular}{llll}
\hline
\hline
exposure & JD  & $t_{\rm exp}$ & seeing  \\
         &     & [s]           & [''] \\
\hline
616247 & 52265.4522 & 70   & 0.61 \\ 
616249 & 52265.4553 & 70   & 0.58 \\
616250 & 52265.4569 & 70   & 0.61 \\
616251 & 52265.4584 & 70   & 0.63 \\
616300 & 52265.5469 & 120  & 1.00 \\
616430 & 52266.3750 & 120  & 0.97 \\
\hline
\hline
\end{tabular}
\begin{footnotesize}
\caption{Basic information about the exposures used
in our analysis. The table gives CFHT exposure number, modified
Julian date, exposure time and seeing. 
\label{tab_data}}
\end{footnotesize}
\end{center}
\end{table}

We analyse the images in exactly the same way as we would in the case
of a weak lensing analysis (e.g., Hoekstra et al. 1998; Hoekstra et
al. 2002a). The only difference is that we measure the shapes of
stars, instead of galaxies. Our shape analysis technique is based on
that developed by Kaiser et al. (1995), with a number of modifications
which are described in Hoekstra et al. (1998) and Hoekstra et
al. (2000). 

We analyse the chips of each exposure separately. After the catalogs
have been corrected for the various observational effects, they are
combined into a master catalog which covers the observed field (for
each pointing). In order to measure the signal on scales out to
one degree we construct a patch of 3 by 5 pointings, which is of a 
size similar to the ones studied by Hoekstra et al. (2002a,2002b) and
van Waerbeke et al. (2002).

\subsection{Shape measurements}

The first step in the analysis is to detect the images of the stars,
for which we used the hierachical peak finding algorithm from Kaiser
et al. (1995). We then select moderately bright stars for which the
shapes are quantified by calculating the weighted central second
moments $I_{ij}$ of the image fluxes and forming the two-component
polarisation

\begin{equation}
e_1=\frac{I_{11}-I_{22}}{I_{11}+I_{22}}~{\rm and}~
e_2=\frac{2I_{12}}{I_{11}+I_{22}}.
\end{equation}

Unweighted second moments cannot be used, because of photon noise.
Instead, a circular Gaussian weight function is used, with a
dispersion equal to the Gaussian scale length of the PSF.  In addition
to the second moments, we compute the shear polarisability tensor
$P^{\rm sh}_{\alpha\beta}$ and the smear polarisability tensor $P^{\rm
sm}_{\alpha\beta}$, which measure the response of an image to a shear
and convolution respectively. Both polarisabilities are computed from
the images themselves. The relevant correct equations can be found in
Hoekstra et al. (1998) (also see Kaiser et al. 1995).

The effect of an anisotropic PSF on the polarisation $e_\alpha$ is
quantified by the smear polarisability. Typically the off-diagonal
terms in the smear polarisability tensor are small, and we therefore
use only the trace in the correction for PSF anisotropy. In this case
the corrected shape of an object is computed using

\begin{equation}
e_\alpha^{\rm cor}=e_\alpha^{\rm obs}-
\frac{P^{\rm obs}_{\alpha\alpha}}{P^\star_{\alpha\alpha}}
e_{\alpha}^\star,
\end{equation}

\noindent where the starred quantities refer to parameters measured
from images of star.

In a typical exposure one can measure the shapes of $\sim 100$ stars
and one needs to interpolate the observed anisotropy in order to
obtain an accurate measure at each position of the chip. Usually,
a second order polynomial 

\begin{equation} 
p_\alpha=a_0+a_1 x +a_2 y + a_3 x^2+ a_4 xy + a_5 y^2,
\end{equation}

\noindent is fitted to the observed values of
$p_\alpha=e_{\alpha}^\star/P^\star_{\alpha\alpha}$ as a function of
position, and this model is used to correct the shapes of galaxies
(e.g., Hoekstra et al. 2002a, 2002b). Alternatively one can fit a
model to $e_{\alpha}^\star(x,y)$ and $P^\star_{\alpha\alpha}(x,y)$
separately (e.g., van Waerbeke et al. 2002). Throughout the paper we
will use the former approach, but we have verified that the latter
method gives similar results.

The number of stars per chip is sufficient to constrain a second order
polynomial, but the use of higher order polynomials can result in
spurious signals (e.g., see the discussion in van Waerbeke et
al. 2002) because the model is poorly constrained at the edges of the
chip. Although a second order polynomial generally fits the
observations well, it is not clear whether it provides the best
description of the data. Hence, a residual PSF anisotropy might be
present due to higher order terms. If this is the case, then how can
one properly account for the higher order terms?

In \S3 we first investigate how well the standard second order model
corrects our data. We then study two strategies where one has detailed
knowledge of the PSF based on observations of fields with many
stars. One approach is to use this ``control'' field (for which we use
exposure 616250) to construct a detailed model of the PSF variation,
and apply this model to observations made at different times. This
strategy was used by Hoekstra et al. (1998, 2000) to correct the
shapes of galaxies in WFPC2 observations. We will refer to this
approach as the ``scaled model'' method. However, the PSF in ground
based data is known to vary more than in space based
observations. Hence we need to examine over what period of time one
can use such a model reliably.  A related approach is to use the star
field to find a suitable parameterization of the PSF variation, and
fit this model to the data (effectively we fit a high order model, but
only include relevant coefficients, resulting in a well constrained
model). This approach is expected to be less sensitive to temporal
variation in the PSF anisotropy, as long as the pattern does not
change completely. As described in \S3, we use a rational function,
and hence we will refer to this strategy as the ``rational function''
method.

\subsection{Ellipticity correlations}

To quantify the effect of imperfect PSF anisotropy correction, we
measure the two ellipticity correlation functions of the residual
shapes. They are given by

\begin{equation}
\xi_{\rm tt}(\theta)=\frac{\sum_{i,j}^{N_s} w_i w_j 
\epsilon_{{\rm t},i}({{\bf x}_i}) \cdot \epsilon_{{\rm t},j}({{\bf x}_j})}
{\sum_{i,j}^{N_s} w_i w_j},
\end{equation}

\noindent and

\begin{equation}
\xi_{\rm rr}(\theta)=\frac{\sum_{i,j}^{N_s} w_i w_j 
\epsilon_{{\rm r},i}({{\bf x}_i}) \cdot \epsilon_{{\rm r},j}({{\bf x}_j})}
{\sum_{i,j}^{N_s} w_i w_j},
\end{equation}

\noindent where $\theta=|{\bf x}_i-{\bf x}_j|$. $\epsilon_{\rm t}$ and
$\epsilon_{\rm r}$ are the tangential and 45 degree rotated shear in
the frame defined by the line connecting the pair of galaxies. The
weights $w_i$ allow for a proper weighting which is needed because of
the noise in the shape measurements. For the following, it is more
useful to consider

\begin{equation}
\xi_+(\theta)=\xi_{\rm tt}(\theta)+\xi_{\rm rr}(\theta),{\rm~and~}
\xi_-(\theta)=\xi_{\rm tt}(\theta)-\xi_{\rm rr}(\theta),
\end{equation}

\noindent i.e., the sum and the difference of the two observed
correlation functions. 

It is important to note that gravitational lensing arises from a
potential and consequently the resulting shear field is curl-free.  As
shown by Crittenden et al. (2002), one can derive ``E'' (curl-free)
and ``B''-mode (pure curl) correlation functions by integrating
$\xi_+(\theta)$ and $\xi_-(\theta)$ with an appropriate window
function. Typically, PSF anisotropy produces both ``E'' and ``B''
modes, and it has been argued that an observed ``B'' mode can be used
as a measure of the systematic signal arising from imperfect
corrections for PSF anisotropy. Also intrinsic alignments of the
sources introduce ``B'' modes (e.g., Crittenden et al. 2002), but this
effect can be removed by using photometric redshift information for
the sources (Heymans \& Heavens 2003; King \& Schneider 2003).

Instead of presenting the ellipticity correlation functions, we
present the results as aperture masses (e.g., Schneider et al. 1998),
as this statistic is commonly used to present the cosmic shear results
(e.g., Hoekstra et al. 2002b; van Waerbeke et al. 2002). Consequently
the results can be compared directly to the published cosmic shear
measurements. The ``E'' and ``B''-mode aperture masses are computed
from the ellipticity correlation functions using

\begin{equation}
\langle M_{\rm ap}^2\rangle(\theta)=\int d\vt~\vt 
\left[{\cal W}(\vt)\xi_+(\vt)+
\tilde{\cal W}(\vt)\xi_-(\vt)\right],
\end{equation}

\noindent and

\begin{equation}
\langle M_\perp^2\rangle(\theta)=\int d\vt~\vt \left[{\cal W}(\vt)\xi_+(\vt)-
\tilde{\cal W}(\vt)\xi_-(\vt)\right],
\end{equation}

\noindent where ${\cal W}(\vt)$, and $\tilde{\cal W}(\vt)$ are given
in Crittenden et al. (2002). Useful analytic expressions were derived
by Schneider et al. (2002). Both ${\cal W}(\vt)$, and $\tilde{\cal
W}(\vt)$ vanish for $\vt>2\theta$, so that $\langle M_{\rm
ap}^2\rangle$ can be obtained directly from the observable ellipticity
correlation functions over a finite interval.

\section{Results}

\begin{figure}
\begin{center}
\leavevmode
\hbox{%
\epsfxsize=\hsize
\epsffile[20 170 520 550]{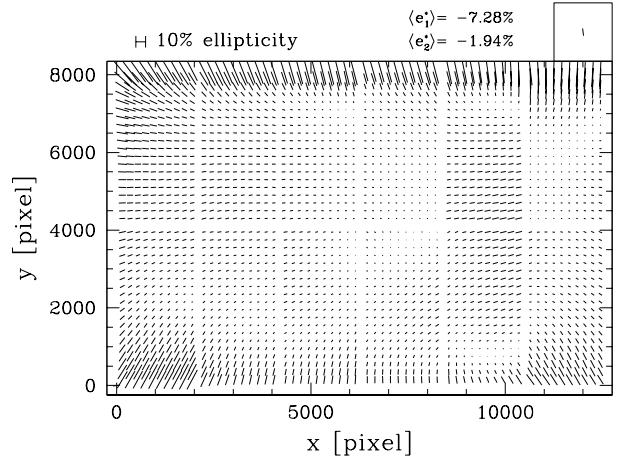}}
\begin{small}      
\caption{PSF anisotropy as a fuction of position for exposure 616250,
which we adopt as our CFH12k reference pointing. The figure shows the
result of a fit to $\sim 2000$ stars on each chip (see \S3 for
details). The PSF anisotropy changes rapidly towards the edges of the
field. We note that this particular pattern is rather extreme. The
sticks indicate the direction of the major axis of the PSF, and the
length is proportional to the observed ellipticity of the PSF. In
order to show the higher order spatial dependence of the anisotropy we
have subtracted the average ellipticity.  The direction of the average
PSF anisotropy is indicated in the top right box, and the amplitude is
indicated as well. Although the PSF anisotropy was determined from
fits to the observed shapes for individual chips, the figure 
shows continuity between chips.
\label{psfmodel}}
\end{small}
\end{center}
\end{figure}

Figure~\ref{psfmodel} shows the result of a detailed model fit to
one of exposure 616250 (see \S3.1 for more details), using all stars
($\sim 2000$ per chip).  To show the higher order spatial dependence
of the anisotropy in more detail, we have subtracted the average
ellipticity.  In the centre of the mosaic the PSF anisotropy does not
vary much, but at the edges the anisotropy increases rapidly to large
values (as much as $20\%$). This pattern is also present in the other
exposures studied in this paper.

The rapid change in PSF anisotropy in the vertical direction is not
well described by the second order polynomial. The pattern seen in the
EXPLORE data could be an extreme, but it could also be a generic
feature of all CFH12k data. We therefore re-inspected the RCS data
(Hoekstra et al. 2002a; 2002b) and searched for the pattern. The PSF
anisotropy is generally small in the RCS data and completely different
from the results shown in Figure~\ref{psfmodel} (see Hoekstra et
al. 2002a). However, inspection of the VIRMOS-DESCARTES data (van
Waerbeke et al. 2002) revealed that the PSF anisotropy pattern in this
case is very similar to the one seen in the EXPLORE data. The origin
of the pattern, and why it is present in the VIRMOS data but not in
RCS, is unclear.

To study the effect of PSF anisotropy on weak lensing measurements, we
``simulate'' a cosmic shear survey. We select about 100 stars at
random on each chip and use these to fit the model for the PSF
anisotropy. The derived model is used to correct the remaining
stars. We repeat this step 15 times. Although the underlying PSF
anisotropy pattern is the same for all 15 samples, the corrections are
slightly different, because different stars were used to derive the
model parameters. The corrected pointings are combined into a large
patch of 3 by 5 pointings, to resemble actual weak lensing
surveys. This patch is used to measure the ellipticity correlation
functions.

\subsection{Comparison of correction schemes}

\begin{figure*}
\begin{center}
\leavevmode
\hbox{%
\epsfxsize=8cm
\epsffile{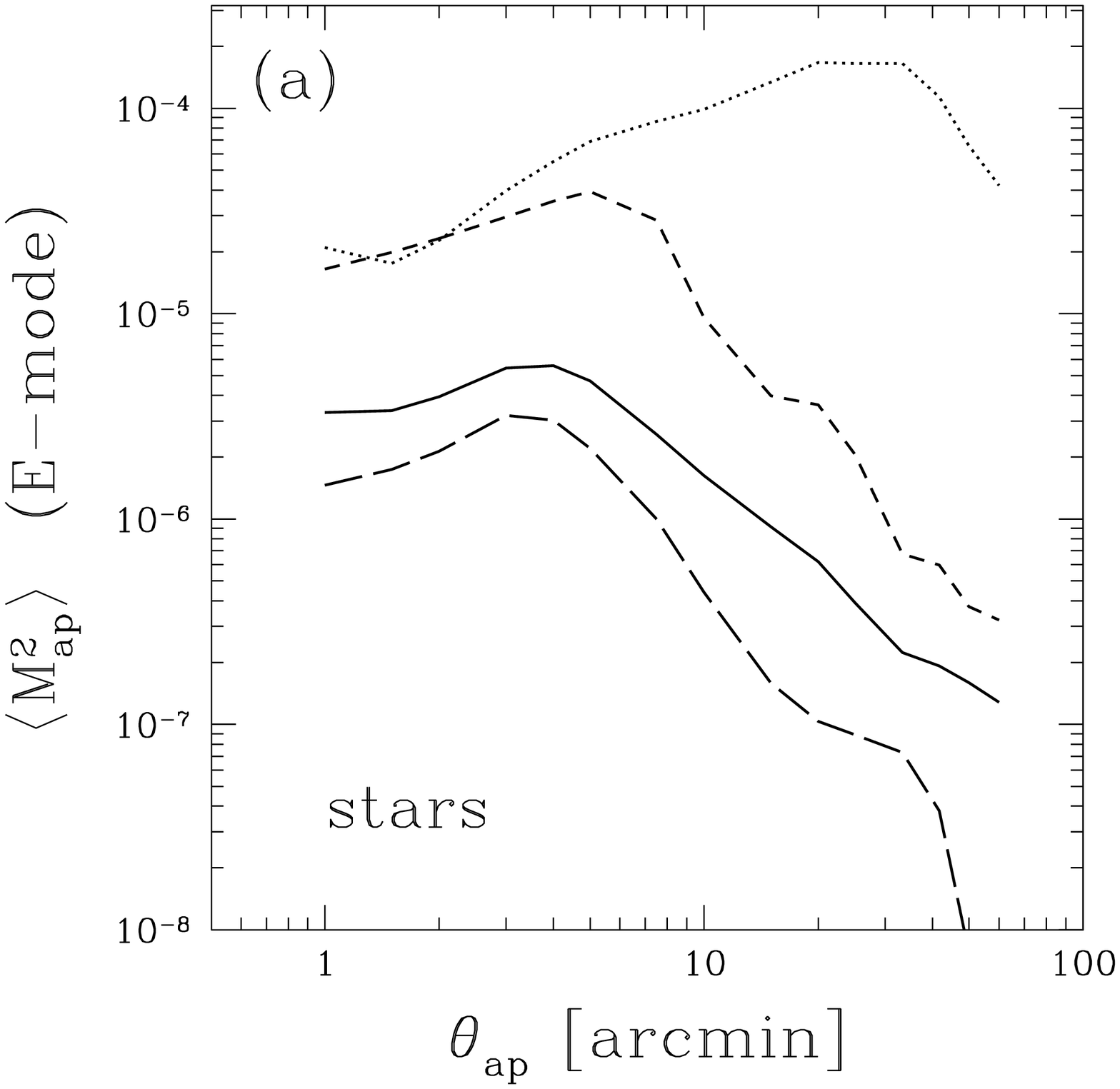}
\epsfxsize=8cm
\epsffile{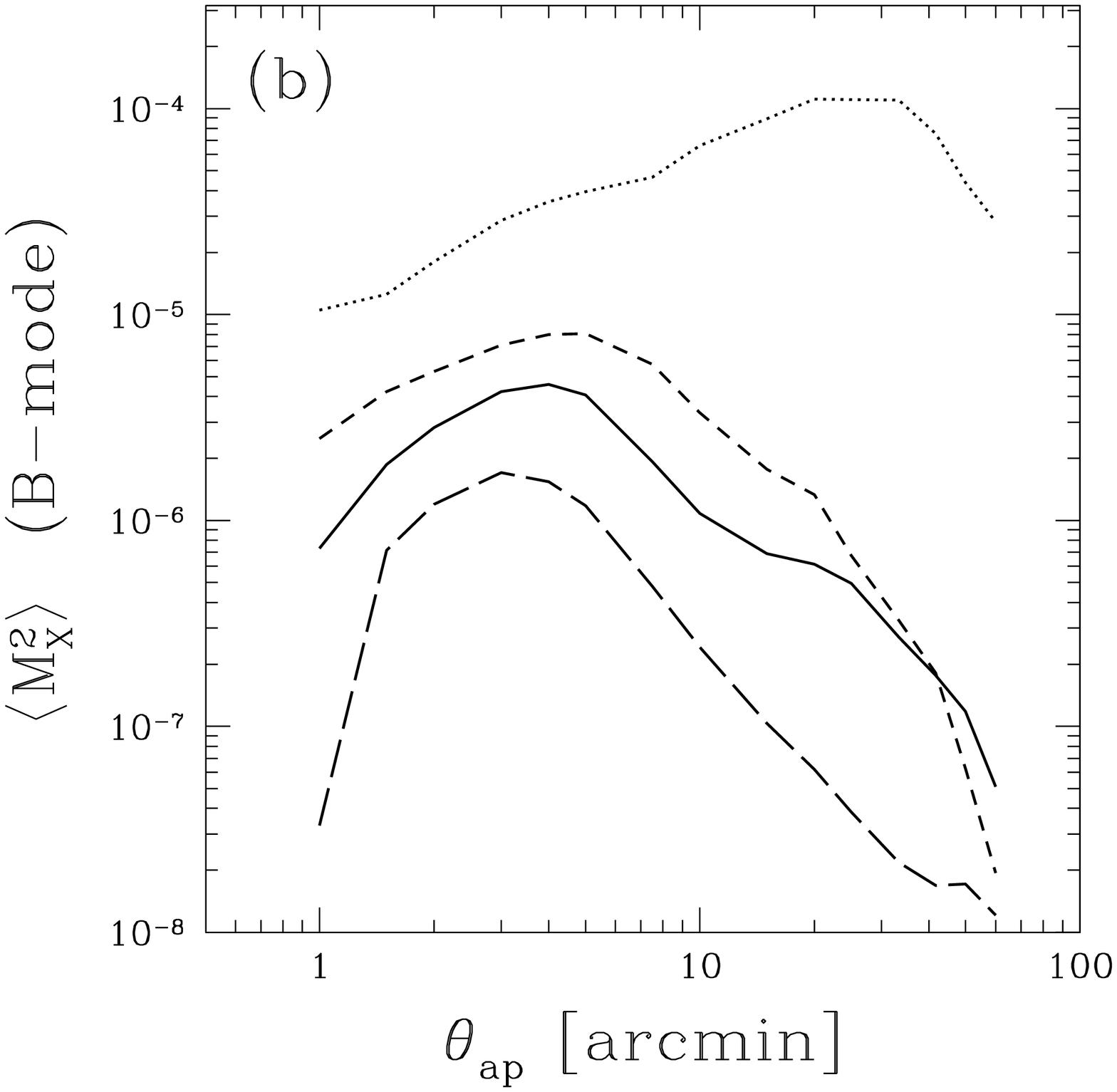}}
\begin{small}      
\caption{(a) The observed amplitude of the variance in the aperture
mass statistic $M_{\rm ap}$ (``E''-mode) as a function of aperture
size. (b) The observed amplitude of the ``B''-mode as a function of
aperture size. The dotted curve indicates the signal
without any correction for PSF anisotropy. The dashed line corresponds
to the results when a second order polynomial is used to characterize
the PSF anisotropy (standard approach). If an additional field with a large
density of stars is observed to obtain an accurate model for the PSF
anisotropy the residual signal can be significantly reduced as is
indicated by the solid line. 
\label{res_psf}}
\end{small}
\end{center}
\end{figure*}

In this section we examine the residual systematics introduced by
various correction schemes. The results presented here use
measurements of exposure 616249. Figure~\ref{res_psf} shows the
resulting signal when Eqn.~3 is used to correct the PSF anisotropy.
The dashed line in panel~a indicates the observed ``E''-mode and
panel~b shows the corresponding ``B''-mode. 

For comparison, we also show the results without PSF anisotropy
correction (dotted lines). The latter results can be compared directly
to Figure~2 from van Waerbeke et al. (2002) who show the aperture mass
variance for the stars in their data. The shape as a function of
aperture size is very similar, but the amplitude in
Figure~\ref{res_psf} is about a factor of 4 higher, which is due to a
combination of different seeing and likely differences in the PSF
anisotropy pattern. In the absence of PSF anisotropy correction, the
maximum signal is reached at an aperture size of $\sim 30$ arcminutes,
which corresponds to a physical scale of $\sim 8$ arcminutes (because
the aperture mass probes smaller scales). This is approximately the
chip scale.

After correction for PSF anisotropy correction the signal peaks at a
physical scale of $\sim 1$ arcminute, which is similar to the average
separation between stars used to derive the model of the PSF
variation. On large scales the standard correction for PSF anisotropy
correction does rather well, because the residuals arise from
imperfect corrections on small scales and the large scale power is
surpressed.

In this particular case the ``B'' mode cannot be used to quantify the
amount of residual systematics in the ``E''-mode: the ``B''-mode is
lower than the ``E''-mode, because the observed signal is caused by a
specific residual pattern. If the residuals were completely random,
one would expect equal ``E'' and ``B'' modes. Obviously the way we
constructed the patches introduces a repeated pattern. We note,
however, that, if the PSF anisotropy is persistent for a reasonable
amount of time, real observations would suffer from a similar problem.
Hence, the observation of a non-zero ``B''-mode is an indicator of
residual systematics, but it does not neccesarily provides a means to
correct the ``E''-mode: in this case subtracting the ``B''-mode from
the ``E''-mode only lowers the ``E''-mode by 30\%.

The large residuals arise because the second order model is a rather
poor fit to this particular PSF anisotropy pattern. The limited number
of stars that can be used in the fit, however, does not warrant higher
order polynomials to be used.

Imperfect guiding of the telescope results in a constant PSF
anisotropy over the whole pointing. This is expected to vary from
exposure to exposure. Higher order terms, however, are likely to be
caused by the telescope optics, and might be relatively stable over a
reasonable period of time. If one were to measure this underlying
``stable'' pattern, one might be able to improve the model for
PSF anisotropy.

For instance we can consider a combination of a second order
polynomial plus a model $c(x,y)$, which is scaled to account for
variations in seeing

\begin{equation} 
p_\alpha=a_0+a_1 x +a_2 y + a_3 x^2+ a_4 xy + a_5 y^2 +a_6 c(x,y).
\end{equation}

The model $c(x,y)$ can be obtained by observing a field with a large
number of stars $(>1000)$ per chip. Such observations allow for a much
more detailed characterization of the variation of the PSF anisotropy.
This approach was used by Hoekstra et al. (1998) to characterize the
PSF anisotropy in WFPC2 observations.

We use all the stars in exposure 616250 to measure the model
$c(x,y)$. We repeat the procedure described above to correct the
shapes of stars in the other exposures, but instead we now use
Eqn.~9. For $c(x,y)$ we adopt the functional form

\begin{eqnarray}
c(x,y)=(c_0+c_1 x +c_2 y + c_3 x^2+ c_4 xy + c_5 y^2+\nonumber\\
c_6 x^3+c_7 y^3 +c_8 y^4)/(1+c_9 x+c_{10}y).
\end{eqnarray}

As mentioned above, the PSF anisotropy changes rapidly towards the
edges of the field, and we found that a rational function provided a
better description compared to a (much) higher order polynomial.  As a
result, the number of parameters used to describe the PSF anisotropy
is still rather low. We note, however, that the model is not a perfect
fit to the data as some very high order residuals are still present.

However, in practice it is not clear how well the ``scaled model''
will work. The pattern might change and one needs to observe fields
with many stars on a regular basis. These fields are likely to be
located in different areas of the sky, and consequently the fact that
the telescope needs to point in a different orientation is likely to
affect the usefulness of the model.

Instead one can use the star fields to find a good parameterization of
the PSF anisotropy variation. High order polynomials require many 
parameters to be fitted, whereas in practice most of the parameters
might have been set to zero. In fact, Eqn.~10 is a much better description
of the data than a fourth order polynomial. Yet, Eqn.~10 requires only
5 more parameters compared to a second order model. 

As an alternative to the ``scaled model'' method, we use Eqn.~10 as
our model for the PSF anisotropy, and fit this model to the data.  The
results are presented in Figure~\ref{res_psf} as solid lines. Although
the residuals are larger than the ``scaled model'', the improvement
over the second order model is substantial. In addition the amplitudes
of the ``E'' and ``B''-modes are similar, and thus one can use the
``B''-mode to correct the lensing signal.

\begin{figure}
\begin{center}
\leavevmode
\hbox{%
\epsfxsize=8cm
\epsffile{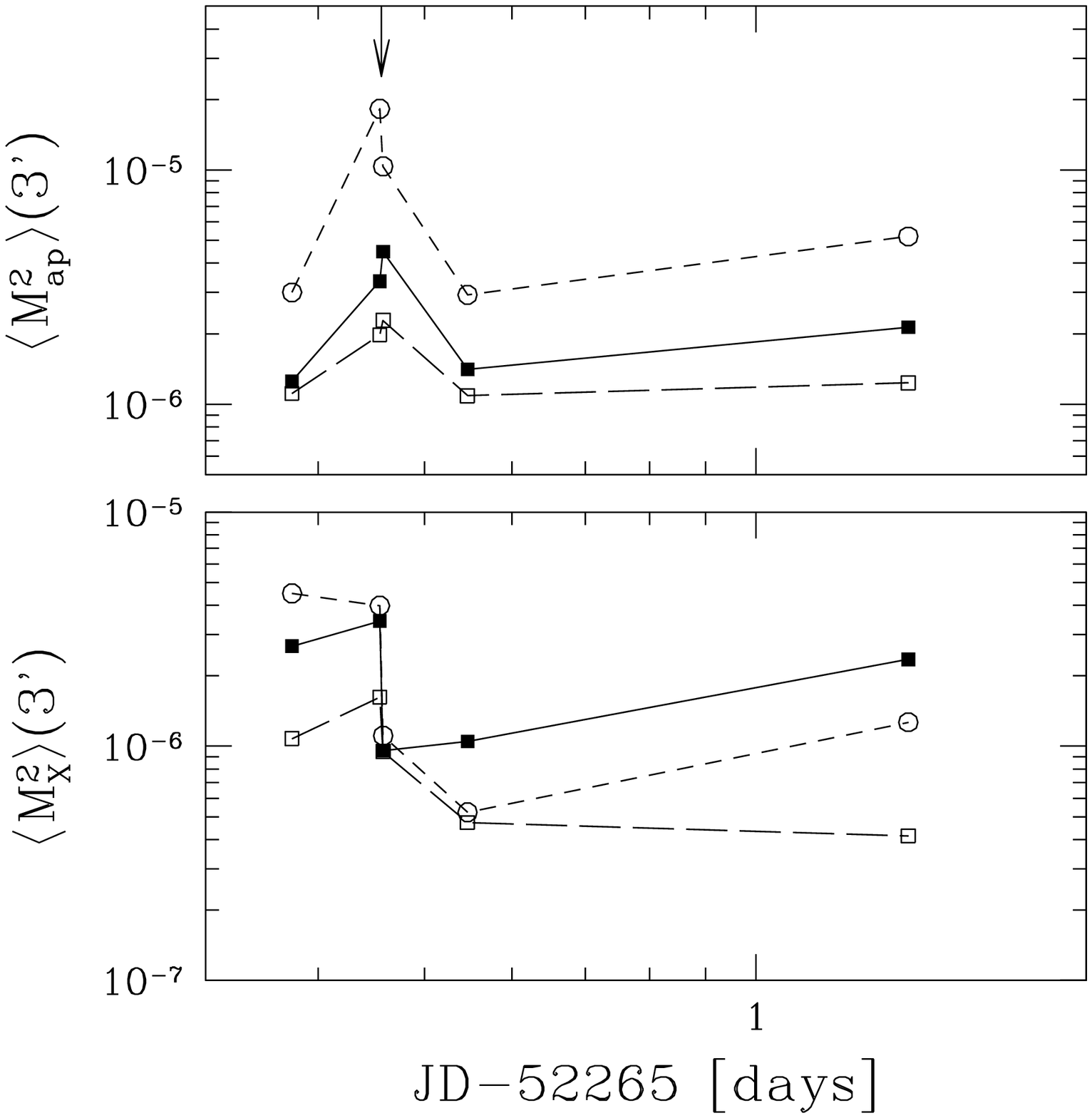}}
\begin{small}      
\caption{The observed amplitude of the variance in the aperture mass
statistic for stars at a scale of 3 arcminutes as a function of
modified Julian date. The upper panel shows the ``E''-mode and the
lower panel the ``B''-mode. The open circles (and dotted lines)
correspond to the 2nd order correction, the open squares (and long
dashed lines) indicate the results for the ``scaled'' model, and the
filled circles (and solid lines) show the results for the rational
function model. The arrow indicates the Julian date of exposure
616250, which was used to derive the ``scaled'' model. The variations
of the signal with time are mostly due to seeing variations. A more
appropriate comparison is discuseed in section 4. Overall, the
``scaled'' model gives the best results, even for observations done a
day later.
\label{mapmax_star}}
\end{small}
\end{center}
\end{figure}

The results of the ``scaled model'' method are indicated by the long
dashed lines in Figure~\ref{res_psf}. Compared to the ``standard''
correction, the residuals are almost an order of magnitude lower. In
addition the ``E'' and ``B''-modes are more similar (but not
identical). 

\subsection{Time variable PSF}

The results presented above suggest that a ``scaled model'' provides
the best correction for PSF anisotropy. In practice, the star field
cannot be observed this close in position and time. We therefore need
to examine whether the ``scaled model'' can be used to correct data
that are taken at different times, with the telescope pointing in
different directions.

The observations listed in Table~\ref{tab_data} span roughly 24 hours,
with the first 4 exposures taken within a few minutes from one
another. Over the period covered by the observations, the orientation
of the telescope changes significantly, as the same field is observed
the whole night.  To examine the time dependence of the correction, we
concentrate on the measurements at a scale of 3 arcminutes, where the
contribution of imperfect PSF anisotropy correction is maximal.

The results are presented in Figure~\ref{mapmax_star} as a function of
modified Julian date. The signal varies significantly, but the
variation is predominantly caused by seeing variation: the signal is
lower when the seeing is larger. A more appropriate comparison is
discussed in \S4.1, where we relate the results to actual weak
lensing measurements. Nevertheless, it is clear that the ``scaled''
model (long dashed lines) gives the best result, even when the model
is applied to data taken the next night.

\begin{figure*}
\begin{center}
\leavevmode
\hbox{%
\epsfxsize=8cm
\epsffile{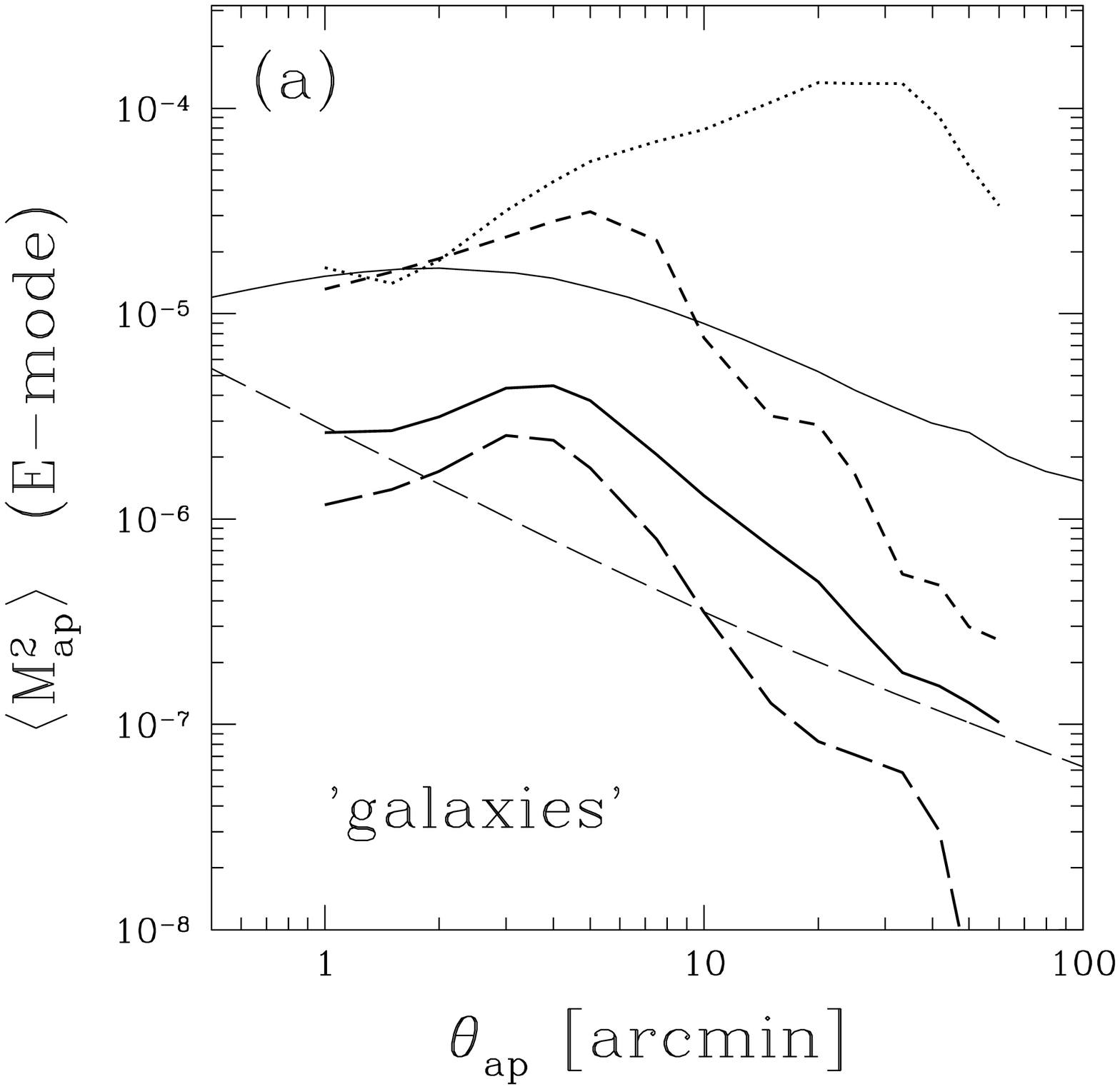}
\epsfxsize=8cm
\epsffile{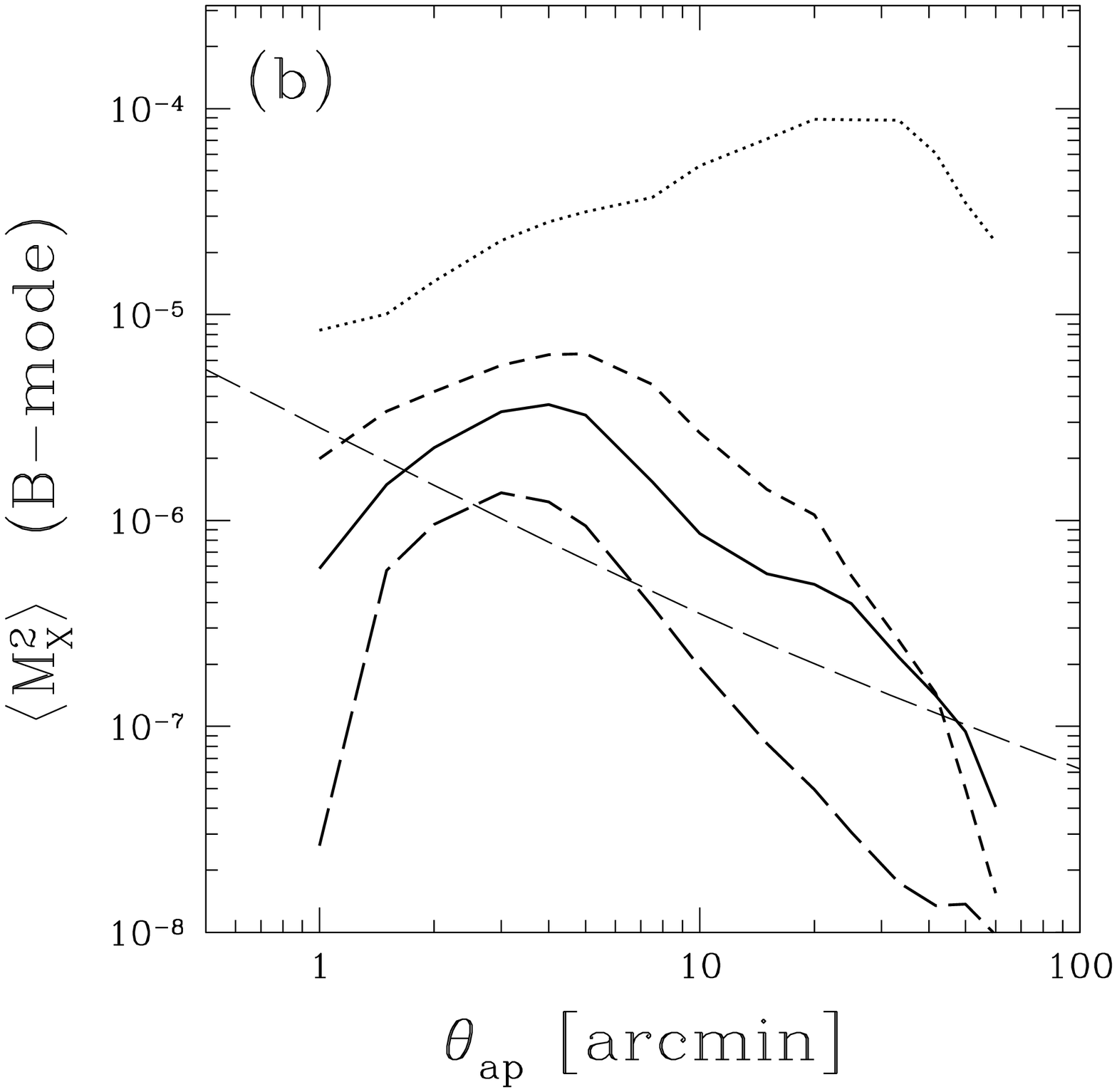}}
\begin{small}      
\caption{(a) The expected amplitude of the variance in the aperture
mass statistic $M_{\rm ap}$ (``E''-mode) as a function of aperture
size in deep observations of galaxies. (b) The expected amplitude of
the ``B''-mode as a function of aperture size.  The dotted curve
indicates the signal without any correction for PSF anisotropy. The
dashed line corresponds to the results when a second order polynomial
is used to characterize the PSF anisotropy (standard approach).  If a
field with a large density of stars is observed to obtain an accurate
model for the PSF anisotropy the residual signal can be significantly
reduced as is indicated by the solid line. The thin solid line
indicates the expected amplitude of the cosmic shear signal from the
CFHTLS and the thin dashed line indicates the projected $1\sigma$
statistical error. The measurements will be dominated by systematics
if the standard correction approach is used, whereas the prospects
are extremely good for the improved scheme.
\label{res_gal}}
\end{small}
\end{center}
\end{figure*}

\section{Effect on cosmic shear studies}

Although it is interesting to study the residual correlations in the
shapes of stars, we would like to know how these results affect actual
cosmic shear measurements. Galaxies have a lower response to PSF
anisotropy than the stars, although the smear polarisabilities are
similar for small, faint galaxies. The correction for the seeing, on
the other hand, will increase the effect of residual PSF anisotropy.
The latter can be very significant for faint galaxies.

It is relatively easy to quantify the effect of imperfect PSF
anisotropy corrections on the shape measurements of galaxies.  We use
values for the smear polarisabilities and seeing corrections using
actual imaging data. We use R-band data from Hoekstra et
al. (2002a). These data were taken using the CFHT, with an integration
time of 15 minutes and seeing ranging from $0\farcs 5-1\farcs0$.

The stars are ``transformed'' into galaxies by changing their
polarisations as

\begin{equation}
e_\alpha^{\rm gal}=\frac{P_{\rm gal}^{\rm sm}}{P_*^{\rm sm}} e_{\alpha}^{*},
\end{equation}

\noindent where the values of $P_{\rm gal}^{\rm sm}$ are drawn from
the RCS imaging data (Hoekstra et al. 2002a). In doing so, we use RCS
exposures matched to the seeing of the EXPLORE data.

We then correct these galaxy shapes for PSF anisotropy using the
different models for the PSF variation. The resulting polarisations
are then corrected for the effect of seeing (e.g., Luppino \& Kaiser
1997; Hoekstra et al. 1998), using the appropriate value of the
pre-seeing shear polarisability $P^\gamma$. We measure the ellipticity
correlation functions and derive the aperture mass statistics.

The results for exposure 616249 are presented in
Figure~\ref{res_gal}. Panel~a shows the curl-free signal, and
Figure~\ref{res_gal}b shows the ``B''-mode.  As expected, the scale
dependence is very similar to the results obtained from the stars, as
is the amplitude of the signal. As before, the thick dotted line
corresponds to the signal without correction for PSF anisotropy
correction, whereas the thick short dashed line indicates the results
for the second order correction. The thick long dashed and solid lines
are for the ``scaled'' model and ``rational function'' model
respectively. Figure~\ref{res_gal} also shows the expected cosmic
shear signal for a $\Lambda$CDM cosmology with $\sigma_8=0.85$ using
the redshift distribution for the sources as given by van Waerbeke et
al. (2002). This signal should be similar to the one we expect to
measure from the CFHTLS.  The thin dashed line indicates the expected
$1\sigma$ statistical error from the CFHTLS based on a scaling of the
errorbars from van Waerbeke et al. (2002) to account for the larger
area of the CFHTLS.

The results obtained here suggest that the use of second order models
can give rise to significant residual systematics. Both Hoekstra et
al. (2002b) and Jarvis et al. (2003) find negligible ``B''-modes on
large scales, but do detect a ``B''-mode on scales smaller than 10
arcminutes. However, the latter two surveys are rather shallow, and
consequently intrinsic alignments are expected to introduce
``B''-modes on scales less than 10 arcminutes.  Hence it is difficult
to separate the contributions arising from both intrinsic alignments
and imperfect PSF anisotropy corrections. 

As mentioned above, the PSF anisotropy pattern seen in the EXPLORE
data is not observed in the RCS data, which typically shows small PSF
anisotropies. Furthermore, the galaxies used in the RCS analysis are
larger than the PSF and as a result the measurements are much less
sensitive to imperfect corrections for the PSF anisotropy (e.g.,
Hoekstra et al. 2002a). These considerations support the conclusion
that the ``B''-mode found by Hoekstra et al. (2002b) is dominated by
intrinsic alignments rather than PSF anisotropy.

The situation is different for the VIRMOS-DESCART survey (van Waerbeke
et al. 2002), which uses fainter, smaller galaxies. Also, examination
of these data show the PSF anisotropy pattern is similar to the one
studied in this paper.  Van Waerbeke et al. (2002) find a small
residual ``B''-mode on scales less than 10 arcminutes, with an
amplitude which is similar to the results presented in
Figure~\ref{res_gal}. Hence, our results suggest that the ``E''-mode
presented by van Waerbeke et al. (2002) might be overestimated on
small scales (also see Figure~10 in van Waerbeke et al.  (2002)). We
have reanalysed the VIRMOS-DESCART data, and found that the improved
correction for PSF anisotropy reduces the small scale variance (on
scales $<10$ arcminutes) by $\sim 30\%$. In addition, the ``B''-mode
after reanalysis is consistent with no signal.

The CFHTLS will be a major improvement over existing surveys in terms
of depth and area. It is reasonable to assume that the PSF anisotropy
of the new Megacam camera will be smaller than that of the CFH12k data
used here. Figure~\ref{res_gal} suggests that the use of appropriate
models for the PSF anisotropy ensures that the systematics are much
smaller than the cosmic shear signal, in particular on large scales.
However, the ultimate goal is to be limited by the statistical errors
only. Although the residuals from the ``rational function''correction
are significant, it is good to note that the amplitudes of the ``E''
and ``B''-modes are similar. Subtracting the ``B''-mode from the
``E''-mode reduces the systematic signal to values below the
statistical noise. Hence, the prospects for accurate cosmic shear
measurements are excellent, provided special attention is paid to
characterizing the variation of the PSF anisotropy.

\begin{figure}
\begin{center}
\leavevmode
\hbox{%
\epsfxsize=8cm
\epsffile{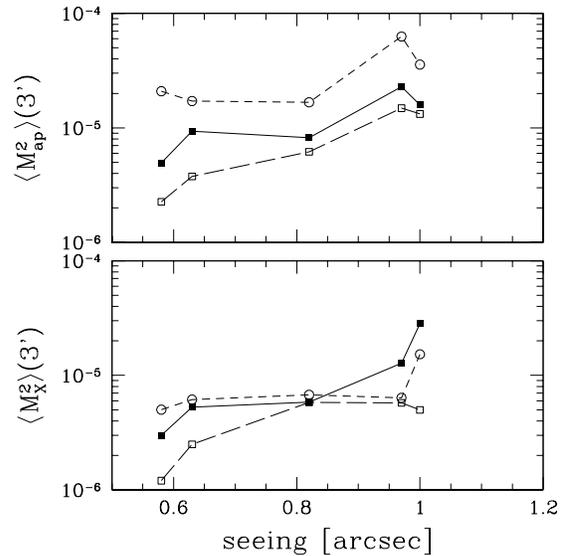}}
\begin{small}     
\caption{The observed amplitude of the variance in the aperture mass
statistic for galaxies at a scale of 3 arcminutes as a function of
seeing. The upper panel shows the ``E''-mode and the
lower panel the ``B''-mode. The open circles (and dotted lines)
correspond to the 2nd order correction, the open squares (and long
dashed lines) indicate the results for the ``scaled'' model, and the
filled circles (and solid lines) show the results for the rational
function model. 
\label{mapmax_gal}}
\end{small}
\end{center}
\end{figure}

\subsection{Effect of seeing}

The EXPLORE data listed in Table~\ref{tab_data} span a range in
seeing, enabling us to examine its effect on the accuracy with which
the weak lensing signal can be measured under different seeing
conditions. The observed PSF anisotropy is smaller when the seeing is
large, but the much larger correction for the circularization by the
PSF will enhance any residual systematics. In this section we examine
whether these two competing effect cancel, or not. We ignore the fact
that the number density of detectable galaxies decreases with
increasing seeing (we effectively assume longer integration times with
increasing seeing).

We compare the amplitudes of the residuals on a scale of 3 arcminutes
as a function of seeing. The results are presented in
Figure~\ref{mapmax_gal}. This figure demonstrates the need of
excellent image quality, as the residual systematics increase with
seeing: poor seeing conditions cannot be fully compensated by taking
longer exposures.

\section{Conclusions}

We have examined the accuracy of the PSF anisotropy correction,
required to measure the lensing signal caused by large scale
structure. We use CFH12k images with a large number of stars which
allow a detailed study of the variation of the PSF with position.

We select $\sim 100$ stars on each chip (similar to the numbers used
in actual weak lensing studies) which are used to derive models for
the PSF anisotropy. We examine three different correction schemes: a
second order polynomial (commonly used in weak lensing studies), a
parameterized model (designed for these particular data), and a
``scaled'' model (derived from a separate field with more than 1000
stars per chip). 

We find that second order models can leave a significant residual
signal. In addition, the ``E'' and ``B''-mode signals are not
identical. Consequently, the observed ``B''-mode cannot be used to
fully correct the cosmic shear signal. Better results are obtained
using an appropriate parameterization of the PSF anisotropy pattern.
In this case we adopt a rational function (Eqn.~10). This approach
reduces the systematics significantly. The best results are obtained
using the ``scaled'' model (Eqn.~9). We find that the pattern is
sufficiently stable in time to warrant the latter approach. 

The PSF anisotropy pattern in the data from the VIRMOS-DESCART survey
(van Waerbeke et al. 2002) is similar to the pattern studied
here. Hence, our results are particularly relevant for this survey,
suggesting that the measurements on scales smaller than 10 arcminutes
are too high. This conclusion is supported by a reanalysis of the
VIRMOS-DESCART data: the small scale variance is reduced by $\sim
30\%$, and the ``B''-mode is consistent with no signal. The improved
VIRMOS-DESCART results are in excellent agreement with the RCS
measurements (Hoekstra et al. 2002b), and the results of the
reanalysis of the VIRMOS data will be published in a forthcoming
paper.

The accuracy with which the cosmic shear signal can be measured
depends critically on the accuracy with which the PSF anisotropy can
be characterized. To ensure minimal contamination of the signal, it is
important that fields with large numbers of stars are observed on a
regular basis. With such an approach it is feasible that large cosmic
shear studies, such as the CFHTLS, will be limited by statistical
noise (caused by the intrinsic shapes of the sources), and not
systematics. In particular measurements on large scales are expected
to be free of systematics. Hence the prospects for high
signal-to-noise measurements of the cosmic shear signal are excellent.

\section*{Acknowledgements}

We would like to thank the EXPLORE Project, in particular Gabriela
Mallen-Ornelas, Howard Yee, Mike Gladders and Sara Seager, for
providing the data which enabled the study presented here. It is
also a pleasure to thank Ludo van Waerbeke for useful discussions.

\end{document}